# Structural Studies of Thin Films of Semiconducting Nanoparticles in Polymer Matrices

*Tiziana Di Luccio,[1,\*] Emanuela Piscopiello,[2] Anna Maria Laera,[2] Marco Vittori Antisari [3]*

[1] ENEA, Centro Ricerche Portici, Via Vecchio Macello, I-80055 Portici (NA), Italy

[2] ENEA, Centro Ricerche Brindisi, SS7 Appia Km 706, I-72100 Brindisi, Italy

[3] ENEA, Centro Ricerche Casaccia, Via Anguillarese 301, I-00060 S. Maria di Galeria (Roma), Italy

Ordered films of nanoscale materials are issue of wide interest for applications in several fields, such as optics, catalysis, bioelectronics. In particular, semiconducting nanoparticles incorporation in a processable polymer film is an easy way to manipulate such materials for their application.

We deposited thin layers of cadmium sulphide (CdS) and zinc sulphide (ZnS) nanoparticles embedded in a thermoplastic cyclo-olephin copolymer (COC) with elevated optical transparency and highly bio-compatible. The nanoparticles were obtained by thiolate precursors previously dispersed in the polymer upon thermal treatment at temperatures ranging between 200 and 300 °C depending on the desired size. The precursor/polymer solutions were spin coated in order to get thin films. The spinning conditions were changed in order to optimise the layer thickness and uniformity. The samples were mainly characterised by X-ray reflectivity (XRR) and by High Resolution Transmission Electron Microscopy (HRTEM) analyses. The thinnest layer we have deposited is 8 nm thick, as evaluated by XRR. The HRTEM measurements showed that the nanoparticles have quasi-spherical shape without evident microstructural defects. The size of the nanoparticles depends on the annealing temperature, e.g. at 232 °C the size of the CdS nanoparticles is about 4 – 5 nm.

---

[\*] Corresponding author.
 *E-mail address*: tiziana.diluccio@portici.enea.it
  Telephone: +39 081 7723244  Fax: +39 081 7723344

# 1. Introduction

The interest in semiconducting nanostructures is due to their size-related physical properties *(quantum size effect)*. The most evident manifestation of such properties is the optical light emission in the blue – red spectral region characterised by a blue-shift at smaller crystallite dimensions [1]. Such properties make semiconducting nanostructures suitable for several kinds of applications, from antireflecting coatings [2] to bioelectronics [3] and light emitting devices [4]. With the advances in organic-based electronics, nanocomposites consisting of nanostructures and polymeric materials have become very attractive for optoelectronic applications [4,5]. Among them, a large effort is devoted to the use of the nanoparticle/polymer nanocomposites as active layer in organic LED (OLED) for new flexible screens in alternative to LCD [6].

Many area of research and industry are investing on polymers since they can be easily manufactured, they are light and optically transparent, and can be processed both in bulk systems and in thin films. At this regard the spin coating technique is widely employed to prepare thin films of polymeric materials [7]. When polymers are used as matrix for nanostructures the arrangement of the embedded nanostructures is influenced by the mass transport properties which in turn depend on the processing temperature. A discontinuity in the viscosity, expected at the glass transition temperature, can deeply influence the particle aggregation in the nanocomposites. In order to overcome such drawback, in-situ growth processes of the nanostructures are employed.

In our work we grow cadmium sulphide (CdS) and zinc sulphide (ZnS) nanoparticles directly within a polymer matrix by a thermolysis process [8,9]. In such a procedure a suitable precursor is dispersed in a polymer solution and a precursor/polymer foil is obtained by casting after the solvent evaporation. Upon heating the precursor/polymer foil the precursor decomposes and the formation of (Cd, Zn)S nanoparticles occurs [9,10]. In the thermolysis method the parameter that mostly influences the size of the nanoparticles is the annealing temperature. Temperatures in the interval (200 – 300) °C produce sizes from 1 to 8 nm [11,12]. In this work we deposit thin films of CdS and ZnS nanoparticles in a COC copolymer by spin coating and successive annealing. The samples produced in this way were studied by High Resolution Transmission Electron Microscopy (HRTEM) and diffraction contrast imaging. Even if nanoparticles contrast is very weak due to their composition as well as the contrast due to the surrounding matrix, TEM images reveal nanoparticles presence in all the prepared samples. X-ray Reflectivity (XRR) was used to investigate the film thickness and surface quality.

2. **Experimental Section**

The samples consisted of CdS and ZnS nanoparticles embedded in thin polymeric films obtained by spin coating on thin microscope cover glass slides (Marienfeld, No.1) for XRR measurements and on NaCl crystals (IR Select, 13 x 2 mm) for TEM measurements. The glass slides were properly cleaned before spinning, by using the RCA protocol [13] and the NaCl crystals were sonicated in acetone and isopropyl alcohol. $Zn(SR)_2$ and $Cd(SR)_2$ bis-thiolates were used as precursors for ZnS and CdS nanoparticles, respectively. The thiolates precursors were prepared by reaction of $Cd(NO_3)_2$ [$Zn(NO_3)_2$] and 2RSH, where R = $-(CH_2)_{11}CH_3$, in ethanol solution [8,9]. Topas (TP) (Ticona, grade 50.13) was used as a polymer matrix for its excellent optical properties (e.g. 93 % light transmission). TP is a thermoplastic cyclo-olephin copolymer constituted of ethylene and norbonene chains. The glass temperature transition $T_g$ and the melting temperature $T_m$ are comprised in the intervals (80 – 180) °C and (240 – 300) °C, respectively. More details on the physical properties of TP can be found in [14] We prepared 1 ml toluene solution for each sample, varying the precursor/polymer w/w % ratio between 10 % and 41 %. First the TP was dissolved in toluene, then the precursor was added to the solution and sonicated until a homogeneous dispersion was obtained. The spin coating was performed by dropping 100 μlt of the above solution on the glass slides and the NaCl substrates rotating at 4000 rpm for 10 seconds. For each value of the concentration we annealed under vacuum one glass slide and one NaCl crystal in the same run in order to impart the same thermal treatment to both samples. All the samples were annealed for 10 minutes at temperature T = 232 °C (except sample S4, annealed at T = 250 °C). All the films are listed in Table 1 where the synthesis parameters are reported.

All TEM images were recorded with a FEI TECNAI G$^2$ F30 transmission electron microscope (ENEA – Brindisi) operating at 300kV with a point-to-point resolution of 0.205nm. The films deposited on NaCl were suitable for TEM observation without any further thinning procedure. After the annealing, the NaCl substrate was dissolved in distilled water, and the TEM specimens were prepared just by collecting the floating film on standard carbon coated TEM grids. Plan-view images of the different samples were recorded at different magnifications.

The XRR measurements were performed by a Philips X'Pert Pro MRD instrument (Univ. of Salerno), equipped with a parabolic multilayer mirror and a four crystal Ge (220) monochromator on the incident beam path, and a parallel plate collimator on the secondary beam path. The x-ray source was a copper sealed tube with $K_{\alpha 1}$ radiation ($\lambda$ = 0.15406nm). Due to the incident beam optics the angular resolution is about 12 arcsec. The reflectivity measurements consisted of coupled $\omega/2\theta$ scans acquired in the reflectivity regime and are represented as a function of the scattering vector $q$ where $q = \dfrac{4\pi \sin(\theta)}{\lambda}$.

## 3. Results

In figures 1a), 1b) and 1c) some examples of the HRTEM images taken from the films S1, S2 and S3 are reported. The length scale is 10 nm in all the images. In the three figures the insets are zooms of the regions included in the white rectangles. The samples S1, S2 and S3 consist of CdS nanoparticles in topas, and they differ from each other for the precursor/polymer ratio: 10 % (S1), 25 % (S2) and 41 % (S3), respectively. In the samples S1 and S2 and in many regions of the sample S3 single particles are quite well separated one from each other, so that they are easily recognisable (dark spots on a brighter background). The average dimensions of the nanoparticles are about 5 nm in S1, 4 nm in S2, between 4 and 10 nm in S3. In all the samples a certain degree of dispersion of both shape and size is observed. Many nanoparticles have an almost spherical geometry in the samples S1 and S2, while an elongated or elliptical shape is dominant in sample S3. By comparing the three samples, S1 and S2 are quite similar from a distribution and size point of view. The particles are more irregular in the sample S3 in terms of average size and size distribution. In the case of ZnS nanoparticles (sample S4) the nanoparticle morphology appears more regular as evident from the HRTEM images in figure 2. Their shape is mainly round, with more homogeneous dispersion respect to the CdS samples. The average size is 4 nm. The HRTEM image of mixed (CdS + ZnS) nanoparticles (sample S5) is reported in figure 3. In this case the estimation of the nanoparticle size is about 5nm, but the size and shape dispersion and the spatial distribution are less regular as evidenced by the white circles and lines in figure 3.

The results of XRR measurements on the films deposited on the glass substrates in the same conditions as the samples S1 and S3 (CdS) are shown in figure 4a and 4b, respectively together with the curve relative to S4 (ZnS) in figure 4c The XRR relative to the sample S5 (ZnS + CdS) is shown in figure 5. We use the same nomenclature used to indicate the corresponding samples on NaCl for the TEM analyses. In figure 4 no thickness fringe is detected in the CdS films, both at low concentration (10.5%, S1) and at higher concentration (41 %, S3). The featureless curves can be due to elevated roughness, as discussed successively. The reflectivity curve relative to the sample S4 consisting of ZnS nanoparticles in TP exhibits a very tiny undulation (see the grey arrows in figure 4c) corresponding roughly to 125 nm as evaluated by the approximate formula for the thickness $t \approx \frac{\lambda}{2\Delta\theta}$ [15]. The surprising results concern the sample S5 formed by CdS and ZnS nanoparticles (figure 5), where a minimum at $q = 0.7$ nm$^{-1}$ is observed, followed by a maximum, typical of a single thickness fringe [15]. The position of the minimum corresponds to a length $d = 9$ nm calculated by taking into account that $d = \frac{2\pi}{q}$. The XRR measurement in figure 5 is reported over a wider angular range respect to figure 4 in order to show the whole oscillation peaked at $q = 1$nm$^{-1}$.

## 4. Discussion

From the HRTEM analyses we can state that films of CdS and ZnS nanoparticles can be prepared by thermolysis of the respective thiolates dispersed in a polymeric matrix and cast as thin film by spin coating of the precursor/polymer solution. At annealing temperatures of 232 °C the size is typically of the order of 5 nm. Figures 1 – 3 indicate that the nanoparticles are predominantly spherical in the case of ZnS, while in the case of CdS almost spherical particles are observed only for a sufficiently small concentration of precursor. The particles are instead elongated or elliptical for the maximum concentration. Respect to CdS nanoparticles grown in bulk polymeric foils previously reported [12] lower annealing temperatures in supported thin films are generally needed to get a given size. For example and as a comparison, 5 nm nanoparticles result from annealing at 280 °C topas and polystyrene foils about 2 mm thick. In terms of growth mechanism, the nanoparticles are formed as a consequence of the thermal decomposition of the starting thiolate precursor [Cd(SR)$_2$ or Zn(SR)$_2$] and successive aggregation of the metal atoms (Cd, Zn) with the sulphur atoms [12]. The polymer matrix has been proved to be inert respect to the nucleation and growth of the nanoparticles as confirmed by wide angle x-ray scattering curves before and after the annealing a[11]. The polymer matrix behaves as viscous fluid able to support the diffusion process required for the particle nucleation and growth, since the working annealing temperatures 232 and 250°C °C are well above $T_g$ of the topas matrix (heat deflection temperature HDT/B = 130 °C) and in the vicinity of the melting point. In the thick polymeric foils, the mobility of the metal and sulphur atoms is three-dimensional while if the precursor molecules are dispersed in thin polymer films the whole growth process is limited to planar regions of nanometric thickness. The speeding up of the synthesis process in samples of reduced thickness can be due to several reasons including a better thermal contact able to easily provide the thermal contribution required by the precursor decomposition process.

Whether the reduced thickness influences the nanoparticle shape and the crystal structure is not yet well understood. In the films studied here the ZnS nanoparticles have clearly a circular projection and are more regularly distributed in the matrix respect to the CdS ones. On the other hand, in the corresponding bulk system the CdS nanoparticles obtained upon annealing the precursor/polymer foil at 276°C are spherical with diameter of about 3nm. Moreover, in the of bulk system while ZnS nanoparticles have a zinc-blende structure independently on the annealing temperature, the CdS nanoparticles undergo a crystal phase transition from zinc-blende to wurtzite phase upon increasing of the annealing temperature [11,12]. Due to the low X-ray diffraction signal produced by the nanoparticles in the thin films samples, we are planning x-ray scattering experiments by means of synchrotron radiation to study the crystal structure and the morphological features of the nanoparticles.

The XRR measurements on the sample S1 and S3 do not show any thickness fringes (figure 4). This can be due to high roughness of the films related to the preparation. The whole growth process of the nanoparticles involves two steps that can easily compromise the thickness uniformity and the overall homogeneity of the films. First, the precursors $Cd(SR)_2$ and $(ZnSR)_2$ are not soluble in common solvents, such as toluene, due to their polymeric structure [16]. Second, as a consequence of such insolubility, the precursor transformation into nanoparticles during the annealing process is influenced by inhomogeneous dispersion of the precursor in the matrix. The solubility might depend on the type of precursor, since the XRR curve of the ZnS sample (S4) shows at least some weak oscillations corresponding to a film thickness of about 125 nm. The unexpected result is the XRR of the mixed (ZnS + CdS) sample (S5), since the single oscillation found at $q = 0.7$ nm$^{-1}$ is indicative of a very thin thickness (9 nm). On the contrary, the expected thickness is about 100 nm, i.e. comparable to the one of sample S4. A parameter that might help to produce more homogeneous films with lower roughness is the solution concentration, i.e. the amount of precursor/polymer ratio respect to the solvent. A first trial at more diluted solutions produced a thin film of about 8 nm thickness as measured by XRR but at these conditions no free standing precursor/polymer film was obtained, suitable for TEM preparation.

## 5. Conclusions

Polymeric films embedding ZnS and CdS nanoparticles were obtained by spin coating and thermolysis of thiolate precursor/polymer solutions. Different concentrations and annealing temperatures were used. HRTEM analyses showed the nanoparticle presence in all the samples, with sizes of about 5 nm at low precursor/polymer concentrations. At higher concentrations, larger particles characterised by size dispersion and aggregation effects were observed. Our current work is devoted to the improvement of the nanoparticles dispersion within the polymer matrix, in order to get higher quality films with low roughness and better thickness uniformity. To this purpose, synchrotron X-ray reflectivity experiments are planned, together with other scattering experiments to investigate the size, shape and distribution in combination with TEM analyses.

The authors thank A. Vecchione of Univ. of Salerno for the use of the X'Pert Pro Diffractometer.

**Tables**

**Table 1.**

List of the precursor/polymer films with the precursor used, precursor/polymer (w/w) concentration (weight expressed in mg), spin coating conditions and annealing temperature. For the spin coating, the solid precursor/polymer dispersion was prepared in 1ml of toluene.

| Sample | Precursor | Concentration precursor/polymer *(w/w)* | Spinning *(rpm /sec)* | Annealing Temperature / Time *( °C / min)* |
|--------|-----------|------------------------------------------|-----------------------|--------------------------------------------|
| S1 | $Cd(SR)_2$ | 9.9 : 39.0 | 4000 / 10 | 232 / 10 |
| S2 | $Cd(SR)_2$ | 19.0 : 46.0 | 4000 / 10 | 232 / 10 |
| S3 | $Cd(SR)_2$ | 4.0 : 38.0 | 4000 / 10 | 232 / 10 |
| S4 | $Zn(SR)_2$ | 11.9 : 36.6 | 4000 / 10 | 250 / 10 |
| S5 | $(Zn + Cd)(SR)_2$ | 17.0 : 45.1 | 4000 / 10 | 232 / 10 |

**List of figure captions**

**Figure 1.** HRTEM images (10 nm scale) of three films of CdS nanoparticles in topas matrix prepared with different precursor/polymer concentrations: 10 %, sample S1 (a), 25 %, sample S2 (b) and 41 %, sample S3 (c). In a) and b) the insets show a zoom of the region included in the white rectangles. The shape is almost circular and the average nanoparticle size is 5 nm and 4 nm, respectively. c) The morphology of the sample S3 is characterised by both spherical and elongated shaped nanoparticles. The nanoparticle distribution is less homogeneous than in S1 and S2. The average size varies between 4 and 10 nm and aggregation is visible in the centre of the figure.

**Figure 2.** HRTEM image of a film of ZnS nanoparticles in topas matrix with precursor/polymer concentration equal to 33 %, sample S4. The nanoparticles are well separated, have circular shape and average size of 4nm. The region in the rectangle is zoomed in the inset.

**Figure 3.** HRTEM image of a film of mixed (CdS + ZnS) nanoparticles in topas matrix with overall precursor/polymer concentration equal to 38 %, sample S5. The zoomed region shows that the nanoparticles have different shapes and different sizes. The average size is about 5nm.

**Figure 4.** X-ray reflectivity curves of two films of CdS, samples S1 (a), S2 (b), and one film of ZnS nanoparticles in topas, S4 (c). While no thickness oscillation is observed in a) and b), due to elevated roughness, some weak oscillations indicated by the arrows are visible in c). The thickness corresponding to the oscillation is 125 nm.

**Figure 5**. X.ray reflectivity curve of a mixed (ZnS + CdS) nanoparticles in topas, sample S5. One single oscillation is detected with its minumum at $q = 0.7$ nm$^{-1}$, typical of a very thin layer (estimated thickness of ~ 9nm).

Figure 1

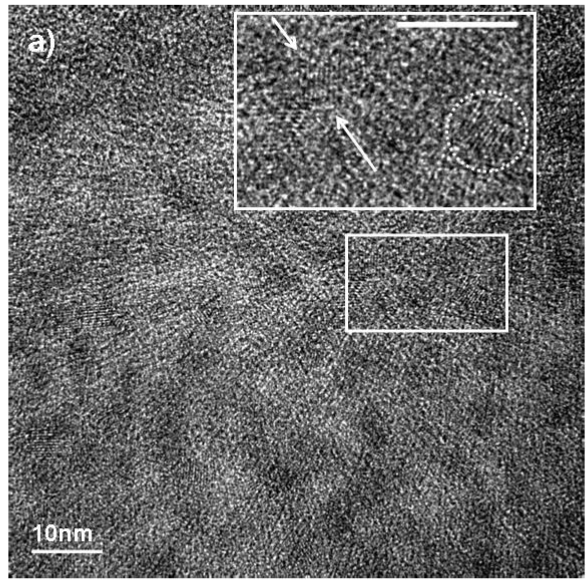

Figure 1

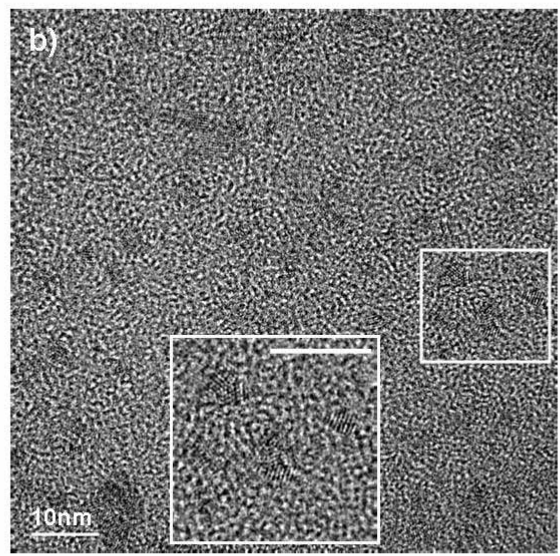

Figure 1

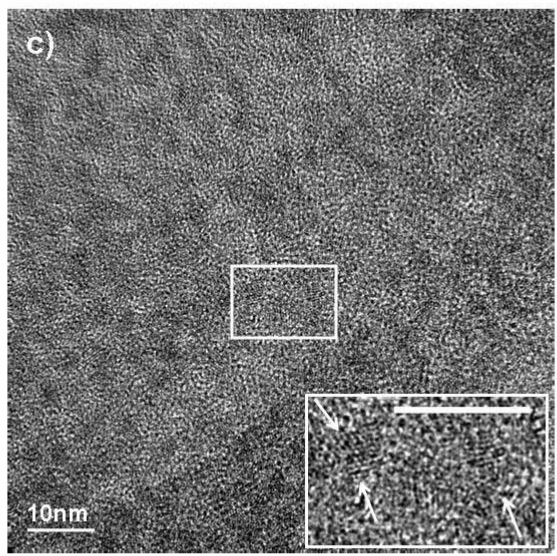

Figure 2

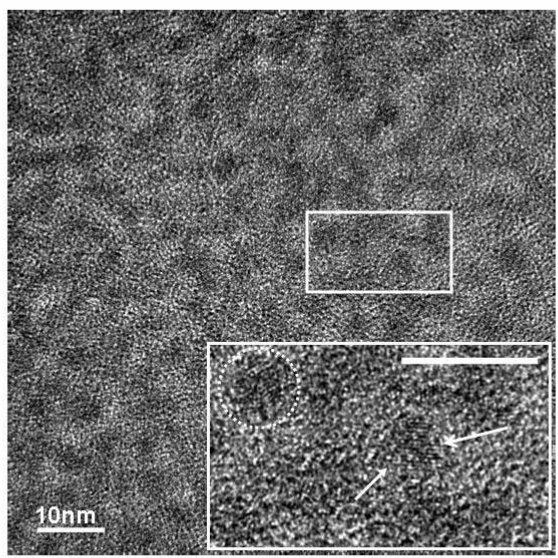

Figure 3

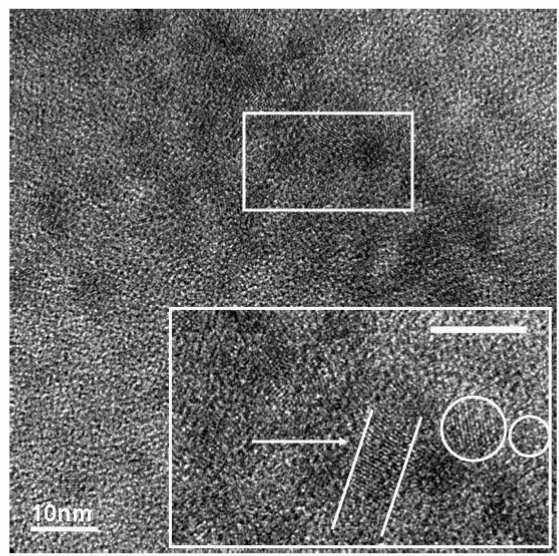

Figure 4

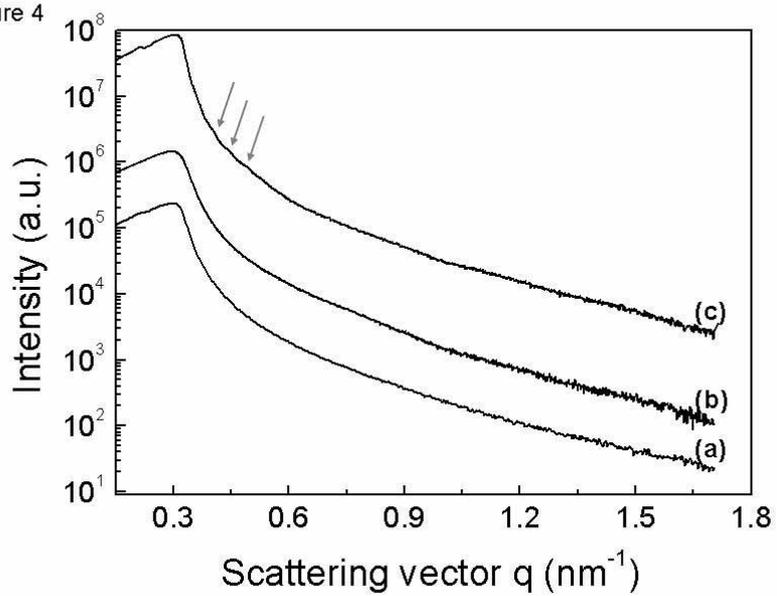

Figure 5

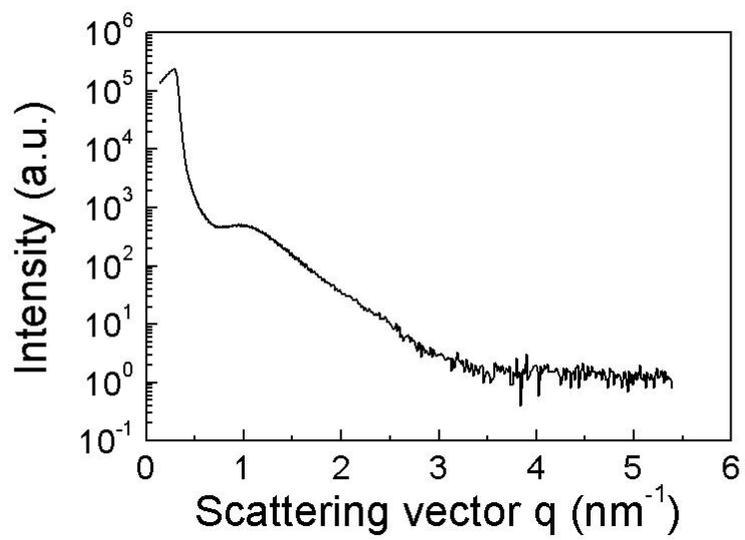